\documentclass[aps,prl,twocolumn,nofootinbib]{revtex4-1}

\usepackage{graphicx}

\begin{document}

\title{Focus Point from Direct Gauge Mediation}

\author{Sibo Zheng}
\affiliation{Department of Physics, Chongqing University, Chongqing 401331, P. R. China}

\begin{abstract}

This paper is devoted to reconcile the tension 
between theoretic expectation from naturalness 
and the present LHC limits on superpartner mass bounds. 
We argue that in SUSY models of direct gauge mediation the focusing phenomenon appears,
which dramatically reduces the fine tuning associated to 126 GeV Higgs boson.
This type of model is highly predictive in mass spectrum,
with multi-TeV third generation, $A_t$ term of order 1 TeV, 
gluino mass above LHC mass bound, 
and light neutralinos and charginos beneath 1 TeV.

\end{abstract}

\maketitle

\section{I.~Introduction}

As the LHC keeps running, 
the searches of supersymmetry (SUSY) signals such as stop/gluino, 
sbottom and Higgs mass discovered at 126 GeV \cite{125} continue to 
push their mass bounds towards to multi-TeV range \cite{ATLAS2013, CMS2013}.
On the other hand, the argument of naturalness requires the masses of third generation scalars, the Higgsinos and gluinos should be $\sim$ 1 TeV.
This is the present status of SUSY.

To reconcile the experimental limits and expectation of naturalness,
either of them needs subtle reconsiderations.
In this paper, we consider relaxing the upper bounds from argument of naturalness.
The upper bounds on above soft breaking parameters arise from the significant contribution to renormalization group (RG) running for up-type Higgs mass squared $m^{2}_{H_{\mu}}$, which connects to the electroweak (EW) scale through electroweak symmetry breaking (EWSB) condition (for $\tan\beta >10$ in the context of the minimal supersymmetric model (MSSM)), 
\begin{eqnarray}{\label{EWSB}}
m^{2}_{Z}\simeq -2\mu^{2}-2m^{2}_{H_{\mu}},
\end{eqnarray}
Naively, low fine tuning implies the value of $\mu$ and $\mid m_{H_{\mu}}\mid$ at EW scale should be both near EW scale.
But there exists an exception.
In some cases, there is significantly cancellation among the RGE corrections arising from soft breaking parameters to $m^{2}_{H_{\mu}}$, 
although their input values are far beyond 1 TeV.
This is known as focusing phenomenon \cite{FPSUSY1, FPSUSY2}.

The early attempts in \cite{FPSUSY1, FPSUSY2, 1201.4338,1303.1622} 
were mainly restricted to SUSY models near grand unification scale (GUT).
One recent work related to focus point SUSY deals with gaugino mediation \cite{gaugino}. 
In this text, we consider gauge mediated (GM) SUSY models with intermediate or low messenger scale $M$ (for a review see, e.g., \cite{Giudice}). 
Since the focusing phenomenon can be analytically estimated only if the gaugino masses dominate over all other soft breaking masses,
or they are small in compared with the third-generation scalar masses (with \cite{FPSUSY3} or without \cite{FPSUSY1, FPSUSY2} $A$ terms ),
following this observation, 
in this paper we study direct GM model,
in which the gaugino masses are naturally small due to the fact that gaugino masses of order $\mathcal{O}(F)$ vanishes \cite{Yanagida}.

Another rational for employing direct GM models is that focusing phenomenon
can be understood as a result of hidden symmetry.
Because without directly gauging global symmetries of the model,
there would be larger symmetries maintained in the hidden theory.
Otherwise, without the protection of symmetry tiny deviation for model parameters from their focus point values leads to significant fine tuning again,
and the model is actually unnatural.

As we will see, there are three free input parameters in our model.
Two of them are fixed so as to induce focusing phenomenon,
leaving an overall mass parameter $m_{0}$.
The fit to 126 GeV Higgs boson discovered at the LHC then determines the magnitude of this parameter, with $m_{0}\sim 4-7$ TeV.
Thus, our model is highly predictive in mass spectrum.

In section IIA, we introduce the model in detail. 
In section IIB, we discuss the focusing phenomenon, 
the boundary conditions for such structure and the mass spectrum at EW scale.
In section IIC, we discuss the possibility of uplifting the gluino mass above LHC lower bound while keeping the focusing.
Finally we conclude in section III.

\section{II.~The Model}
\subsection{A.~Setup}
In contrast to \cite{Agashe}, in which non-minimal GM model was employed to discuss focusing phenomenon, 
we study SUSY models that don't spoil the grand unification of SM gauge couplings 
and restrict to the context of direct GM.
The messenger fields include chiral quark superfields $q+q'$ and their bi-fundamental fields $\bar{q}+\bar{q'}$, 
lepton superfields $l+l'$ and their bi-fundamental fields $\bar{l}+\bar{l'}$, and singlet $S$ and its bi-fundamental field $\bar{S}$.
They transform under $SU(3)_{C} \times SU(2)_{L}\times U(1)_{Y}$ as, respectively, 
\begin{eqnarray}{\label{model1}}
q, ~q' &\sim& \left(\mathbf{3}, \mathbf{1}, -\frac{1}{3}\right),\nonumber\\
\bar{q},~\bar{q'}&\sim& \left(\bar{\mathbf{3}}, \mathbf{1}, \frac{1}{3}\right),\nonumber\\
l, l' &\sim& \left(\mathbf{1}, \mathbf{2}, \frac{1}{2}\right),\\
\bar{l},~\bar{l'} &\sim& \left(\mathbf{1}, \bar{\mathbf{2}}, -\frac{1}{2}\right)\nonumber\\
S,~\bar{S} &\sim& \left(\mathbf{1}, \mathbf{1}, 0\right)\nonumber
\end{eqnarray}
So, these messenger multiplets complete a $\mathbf{5}+\bar{\mathbf{5}}$ representation of SM gauge group.
The renormalizable superpotential consistent with SM gauge symmetry is given by \footnote{It belongs to general Wess-Zumino model, 
which can be completed as effective theory of strong dynamics at low energy \cite{ISS}. 
The direct gauge mediation arises after gauging the global symmetries in the weak  theory and identifying them as SM gauge groups.},
\begin{eqnarray}{\label{potential}}
W=fX+Xq\bar{q}+Xl\bar{l}+m(q'\bar{q}+q\bar{q'})+m(l'\bar{l}+l\bar{l'}).\nonumber\\
\end{eqnarray}
where $X=M+F\theta^{2}$, denotes the SUSY-breaking sector with nonzero $F$ term.
In what follows, we will consider $N$ copies of such messengers multiplets,
with $N<6$ so as to maintain the grand unification of SM gauge couplings.

For the purpose of focusing we add a deformation to superpotential Eq.(\ref{potential}),
\begin{eqnarray}{\label{deformation}}
W=\lambda H_{u}S\bar{l}.
\end{eqnarray}
This  superpotential can be argued to be natural by either imposing a hidden $U(1)_X$ symmetry \cite{soft2} or matter parity \cite{1007.3323}.
For example, we can impose $U(1)_X$ charges of fields as, 
\begin{eqnarray}{\label{parity}}
q_{X}(X,~\phi_{i},~\bar{\phi}_{i},~H_{u}, H_{d})=(1,-1/2,-1/2,1,-1)
\end{eqnarray}
where $\phi_{i}=\{q,q', l, l' , S\}$.
In addition, this hidden symmetry forbids some operators such as $H_{d} Sl$.

In Eq.(\ref{potential}) we have assumed unified mass parameter $m$ and ignored the Yukawa coefficients for simplicity.
For $m< M$ which we adopt in this paper
the soft scalar mass spectrum induced by superpotential Eq.(\ref{potential}) is the same as that of minimal GM
at the leading order.
Since the minimal GM can not induce focusing phenomenon,
the deformation to the scalar mass spectrum due to Eq.(\ref{deformation}) is thus crucial for our purpose.
In particular, Eq.(\ref{deformation}) gives rise to a negative one-loop contribution to $m^{2}_{H_{u}}$ with suppression factor $F/M^{2}$. 
Unless we take $\sqrt{F}<< M$, the sign of $m^{2}_{H_{u}}$ would be negative,
it will not lead to focusing (see explanation around Eq.(\ref{mass2})).
Therefore, we are restricted to choose 
\begin{eqnarray}{\label{limit}}
m< M,~~\text{and}~~\sqrt{F}<< M.
\end{eqnarray}
For detailed calculation of the deviation to scalar mass spectrum given by Eq.(\ref{deformation}),
We refer the reader to \cite{soft1, soft2}.
With small SUSY breaking given by Eq.(\ref{limit}), 
$m^{2}_{H_{\mu}}$ will be uplifted as required for focusing.

One can verify that gaugino masses at one loop of order $\mathcal{O}(F)$ vanish 
due to the fact the mass matrix of messengers 
\begin{eqnarray}{\label{matrix}}
\mathcal{M}=\left(
\begin{array}{ll}
X & m \\
m & 0 
\end{array}\right)
\end{eqnarray}
satisfies $\det\mathcal{M}=\text{const}$ as long as $m$ doesn't vanish,
although $m$ is small in comparison with scale $M$. 
So we expect that the RGE for $m^{2}_{H_{\mu}}$ is dominated by stop mass squared $m^{2}_{Q_{3}}$, $m^{2}_{u_{3}}$, and Eq.(\ref{deformation}) induced A term.

\subsection{B.~Focusing And Mass Spectrum}
\begin{figure}[ht]
\includegraphics[width=0.45\textwidth]{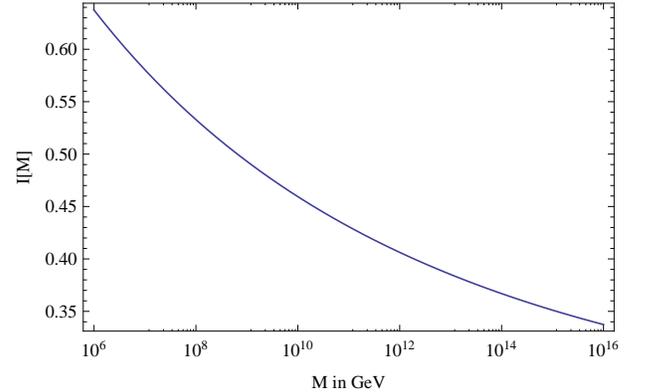}
\caption{$I$ as function of $M$ for the context of MSSM below scale $M$. }
\label{I}
\end{figure}

Following the observation \cite{FPSUSY1, FPSUSY2, FPSUSY3} that the REGs for $A_t$ and scalar masses such as $m^{2}_{H_{\mu}}$ are affected by both themselves and gluino masses,
while the RGE for gluino mass is only affected by itself,
we can solve the RGEs for soft scalar masses, 
\begin{eqnarray}{\label{rge}}
\left(%
\begin{array}{c}
  m^{2}_{H_{\mu}} (Q)\\
  m^{2}_{u_{3}} (Q) \\
  m^{2}_{Q_{3}} (Q)\\
  A^{2}_{t} (Q) \\
\end{array}%
\right)&=&\kappa_{12}I^{2}(Q)
\left(%
\begin{array}{ccccc}
  3   \\
  2  \\
  1  \\
  6  \\
\end{array}%
\right)+
\kappa_{6}I(Q)
\left(%
\begin{array}{ccccc}
  3   \\
  2  \\
  1  \\
  0  \\
\end{array}%
\right)\nonumber\\
&+&
\kappa_{0}
\left(%
\begin{array}{ccccc}
  1   \\
  0  \\
  -1  \\
  0  \\
\end{array}%
\right)+
\kappa'_{0}
\left(%
\begin{array}{ccccc}
  0  \\
  1  \\
  -1  \\
  0  \\
\end{array}%
\right).
\end{eqnarray}
for small gluino masses (in compared with above scalar soft masses).
Here, 
\begin{eqnarray}{\label{zeta}}
I(Q)=\exp\left(\int^{\ln Q}_{\ln M} \frac{6y^{2}_{t}(Q')}{8\pi^{2}} d\ln Q'\right)
\end{eqnarray}
which depends on $M$ and RGE for top Yukawa.
In Fig. \ref{I} we show the numerical value of $I$ as function of $M$, with the context of MSSM below scale $M$.
In particular, $I(1 \text{TeV})\simeq 0.527$ for $M=10^{8}$ GeV.

The condition for focusing phenomenon can be derived from Eq.(\ref{rge}) by imposing  $m^{2}_{H_{\mu}}(1 \text{ TeV})\simeq 0$.
Define $m^{2}_{H_{\mu}}(M)=+m^{2}_{0}$.
Similar to \cite{FPSUSY3} we choose $x$ to parameterize the splitting between $m^{2}_{Q_{3}}(M)$ and $m^{2}_{u_{3}}(M)$, and $y$ to be directly related to $A_{t}(M)$. 
In the case of small SUSY breaking the mass spectrum which induces focusing at scale $\mu=1 \text{TeV}$ reads as,
\begin{eqnarray}{\label{mass}}
m^{2}_{0}\left(%
\begin{array}{c}
  1\\
  1.41+x-1.58y \\
  1.82-x-3.16y\\
  9y\\
\end{array}%
\right)_{M}\rightarrow
m^{2}_{0}\left(%
\begin{array}{c}
  0\\
  0.74+x-1.58y\\
  1.48-x-3.16y\\
  1.66y\\
\end{array}%
\right)_{\mu}\nonumber\\
\end{eqnarray}

Alternatively we rescale parameter $x$ as in \cite{FPSUSY3} such that $m^{2}_{Q_{3}}$ only depends on $x$.
For $m^{2}_{H_{\mu}}(M)=-m^{2}_{0}$, Eq.(\ref{mass}) is instead of, 
\begin{eqnarray}{\label{mass2}}
m^{2}_{0}\left(%
\begin{array}{c}
  -1\\
  -1.41+x-1.58y \\
  -1.82-x-3.16y\\
  9y\\
\end{array}%
\right)_{M}\rightarrow
m^{2}_{0}\left(%
\begin{array}{c}
 0\\
  -0.74+x-1.58y\\
  -1.48-x-3.16y\\
  1.66y\\
\end{array}%
\right)\nonumber
\end{eqnarray}
This parameterization appears when $F/M^{2}\rightarrow 1$.
In this limit, $m^{2}_{H_{\mu}}$ is dominated by the one-loop negative contribution  proportional to Yukawa coupling $\lambda$. 
From Eq.(\ref{mass2}), 
there is no consistent solution to $x$ and $y$ in this case.

Soft masses in Eq.(\ref{rge}) at scale $\mu =1 \text{TeV} $ are functions of Yukawa coupling $\lambda$, 
number of messenger pairs $N$, ratio $F/M^{2}$ and SUSY-breaking mediated scale $M$.
From Eq. (\ref{mass})  one connects the variables $(x, y)$ and the model parameters $\lambda$ and $N$.
For the three input parameters $m_{0}$, $x$ and $y$ (with $M$ fixed) for focusing in the model,
two of them can be fixed by the choices of $\lambda$ and $N$. We choose $x$ and $y$ for analysis.
Fig.\ref{xy} shows the plots of $x$ (dotted) and $y$ (solid) as function of $\alpha_{\lambda}$ and $N$. 
For each $N$ the focus point values of $x$ and $y$ are read 
from the crossing points between vertical line and solid curve (dotted curve ) for $y$ ($x$) .
Therefore, there is only one free parameter left in the model by imposing the focusing condition,
which is very predictive in the mass spectrum and signal analysis.

Since we perform our analysis in perturbative theory,
in order to avoid Landau pole up to GUT scale, 
the Yukawa coupling $\alpha_{\lambda}$ is upper bounded, $\sim$0.1 for our choice of messenger scale. 
The dotted and solid horizontal lines in fig. \ref{xy} refer to allowed ranges for $x$ and $y$, respectively. These ranges are derived from the requirement that the stop soft masses aren't tachyon-like and the $A_t$ squared is positive.
Following these we obtain,
\begin{eqnarray}{\label{ranges}}
0 <y < 0.40,~-0.74<x<1.48,\nonumber\\
1.58 y-0.74<x<1.48-3.16y,\\
1.58 y-1.41<x<1.82-3.16y.\nonumber
\end{eqnarray}
It is easy to verify that for each $N$ the crossing points  satisfy the constraints above.
\begin{widetext}
\begin{center}
\begin{figure}[ht]
\includegraphics[width=0.7\textwidth]{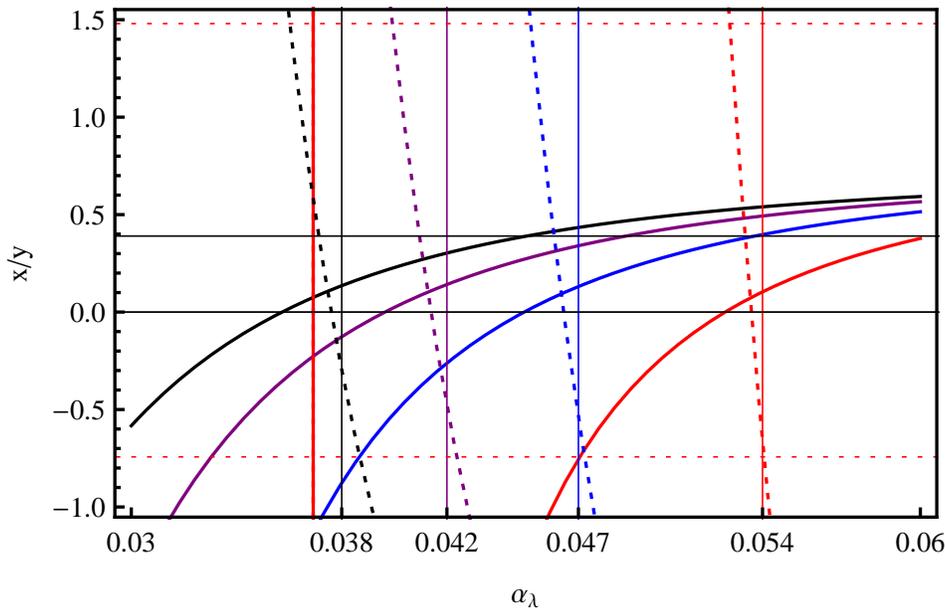}
\caption{Plots of $x$ (dotted) and $y$ (solid) as function of $\alpha_{\lambda}$ for $N=\{1,2,3,4\}$.  The red, blue, purple and black curves correspond to $N=1,2,3,4$, respectively. For each $N$ the focus point value are read from the crossing points between vertical line and solid curve for $y$ and dottoed curve for $x$, respectively. The dotted (solid) horizontal lines refer to range allowed for $x$ ($y$).}
\label{xy}
\end{figure}
\end{center}
\end{widetext}

\begin{figure}[ht]
\includegraphics[width=0.45\textwidth]{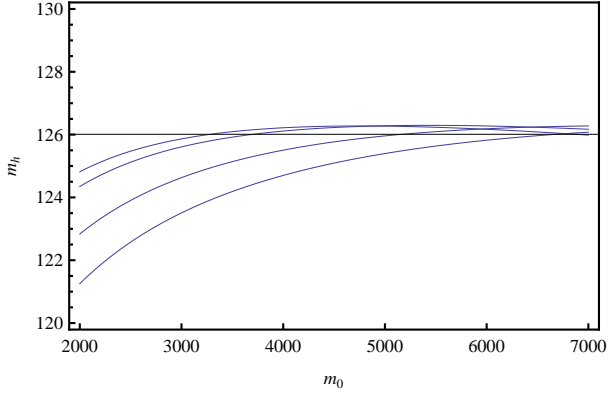}
\caption{$m_{h}$ vs $m_{0}$ for different $N$s, 
with $N=1,2,3, 4$ from bottom  to top, respectively. 
Multi-TeV $m_{0}$ is required by the 126 GeV Higgs boson.}
\label{m0}
\end{figure}

With focusing phenomenon we have single free parameter, namely $m_{0}$ at hand.
It can be uniquely determined in terms of the mass of Higgs boson observed at the LHC.
Fig. 2 shows how $m_{h}$ changes as parameter $m_{0}$ for different $N$s. 
The two-loop level Higgs boson mass in the MSSM is given by \cite{Carena},
\begin{widetext}
\begin{eqnarray}{\label{b}}
m_{h}^{2}&=& m_{Z}^{2}\cos^{2}2\beta+\frac{3m^{4}_{t}}{4\pi^{2}\upsilon^{2}}\left\{\log\left(\frac{M^{2}_{S}}{m^{2}_{t}}\right)
+\frac{1}{2}\tilde{A}_{t}
+\frac{1}{16\pi^{2}}\left(\frac{3}{2}\frac{m^{2}_{t}}{\upsilon^{2}}-32\pi \alpha_{3}\right)\left[\tilde{A}_{t}
+\log\left(\frac{M^{2}_{S}}{m^{2}_{t}}\right)\right]\log\left(\frac{M^{2}_{S}}{m^{2}_{t}}\right)\right\}
\end{eqnarray}
\end{widetext}
Here $\upsilon=174$ GeV and $\tilde{A}_{t}=\frac{2X^{2}_{t}}{M^{2}_{S}}\left(1-\frac{X^{2}_{t}}{12M^{2}_{S}}\right)$, with $X_{t}=A_{t}-\mu\cot\beta$.
We focus on large $\tan\beta$ region.
For  $\tan\beta \geq 20$, 
the fit to Higgs boson mass doesn't change much.
From fig.\ref{m0} one observes that $m_{0}\sim4.0-7.0$ due to the fit to 126 GeV Higgs boson.
Substituting the values of $m_{0}$ from fig.\ref{m0} and $x$, $y$ from fig.\ref{xy} into Eq.(\ref{mass})
we find the mass spectrum,
which is shown in table 1.

The choice on large $\tan\beta$ might be forbidden by possibly large B$\mu$ term induced by Eq.(\ref{deformation}). As noted in \cite{1007.3323},  B$\mu\sim \mu A_{t}$. 
In terms of electroweak symmetry breaking condition, we have $\sin(2\beta)\simeq \text{B}\mu/m^{2}_{0}\sim (A_{t}/m_{0})^{2}\cdot (\mu/A_{t})$. 
With a small $\mu$ term of order $\sim 300-500$ GeV (as shown in table 1) at messenger scale $M$, 
one does not have to worry about $\mu$ being made very large by radiative correction involving heavy soft scalar masses (see e.g., \cite{soft1}).
So, one obtains $\sin(2\beta)$ of order $\sim (1/4)^{2}\cdot (1/4)$ from table 1,
and the choice on large value of $\tan\beta$ is not violated by B$\mu$ term.

\subsection{C.~Gaugino Mass}
As mentioned above
due to $\det\mathcal{M}=\text{const}$ gaugino masses vanish 
at one-loop level of order $\mathcal{O}(F)$ and at the two-loop level of order $\mathcal{O}(F)$.
Their leading contributions appear at one-loop level of order $\mathcal{O}(F^{3}/M^{5})$ \cite{Yanagida}.
Under small SUSY-breaking limit 
 the magnitude of gaugino mass relative to $m_{Q_{3}}$ at input scale is given by 
\footnote{We thank the referee for pointing out a critical error in estimation of gaugino mass in the previous version of this manuscript.},
\begin{eqnarray}{\label{gaugino}}
\frac{m_{\tilde{g}_{i}}}{m_{Q_{3}}}\sim\left(\frac{F}{M^{2}}\right)^{2}\cdot\frac{\sqrt{N}\alpha_{i}}{\sqrt{2\times \left(\frac{4}{3}\alpha^{2}_{3}(M)+\frac{3}{4}\alpha^{2}_{2}(M)+\frac{1}{60}\alpha^{2}_{1}(M)\right)}}\nonumber\\
\end{eqnarray}
Using one-loop RGEs for gluino masses,
we find their values at the renormalization scale $\mu=1$ TeV.
One observes from Eq.(\ref{gaugino})
that the gluino mass is far below the 2013 LHC bound $\simeq 1.3$ TeV due to the suppression by factor $F^{2}/M^{4}$.

Without extra significant modifications to the gaugino mass spectrum,
LHC bound would exclude this simple model,
despite it provides a natural explanation of Higgs boson mass 
and is consistent with present experimental limits.
Here, we propose a recipe \cite{0612139} in terms of imposing small modification to 
superpotential $\delta W =m' \left( \bar{l'}l'+\bar{q'} q'\right )$, with small mass $m' <m$.
These mass terms are consistent with gauge symmetries and matter parity of messenger sector. 

If so, Eq.(\ref{matrix}) will be instead of
\begin{eqnarray}{\label{matrix2}}
\mathcal{M}=\left(
\begin{array}{ll}
X & m \\
m & m' 
\end{array}\right)
\end{eqnarray}
The correction to soft scalar mass spectrum is of order $\mathcal{O}(m'^{4}/m^{4})$ and very weak.
However, the correction to gaugino mass, which is of order,
\begin{eqnarray}{\label{gaugino2}}
m_{\tilde{g}_{i}}\simeq N \cdot\frac{\alpha_{i}}{4\pi}\cdot\frac{F}{m}\cdot\frac{m'}{m}
\end{eqnarray}
can be large enough to reconcile with the LHC bound when $m'/m$ is larger than $F^{2}/M^{4}$.
For example,  we choose $N=1$,  $M=10^{8}$ GeV and $m=0.1 M$.
Then $m_{0}\sim 7$ TeV and $\sqrt{F}\sim 8.2 \cdot 10^{6} $ GeV, 
and further $m_{\tilde{g}_{3}}\sim 7 \cdot 10^{-3} \cdot m'$ from Eq.(\ref{gaugino2}).
LHC gluino mass bound requires $m' \geq 2 \cdot 10^{5}$ GeV,
which is consistent with the constraint $m' <m<M$.
The bino and wino masses are both near 1 TeV.
So they are the main target of 14-TeV LHC.

\begin{table}
\begin{center}
\begin{tabular}{|c|c|c|c|c|}
  \hline
   & $N=1$ & $N=2$ & $N=3$& $N=4$\\
  \hline
  $m_{0}$ & $7.0$ & $5.9$ & $4.0$ & $3.5$\\
 \hline
  $m_{\tilde{t}_{1}}$ & $3.12$ & $3.62$ & $4.54$ & 4.83\\
  $m_{\tilde{t}_{2}}$ & $7.65$ & $4.98$ & $4.80$& 6.0 \\
  $A_{t}$ & $1.64$ & $1.48$ & $1.50$& 1.50 \\
  $\mu$ & $0.50$ & $0.42$ & $0.28$& 0.24\\
  \hline
\end{tabular}
\caption{Given a focus point, 
input mass parameter $m_{0}$ (in unit of TeV) required for $m_{h}=126$ GeV and corresponding soft mass spectrum (in unit of TeV) at renormalization scale $\mu=1$ TeV in the context of MSSM, for different values of messenger number $N$.}
\end{center}
\end{table}

\section{III.~Discussion}
From mass spectrum of table 1,
the main source for fine tuning arises from $\mu$ term.
The fine tuning parameter c,
which is defined as $c=\max\{c_{i}\}$, with
\begin{eqnarray}
 c_{i}=\mid \partial \ln m^{2}_{Z}/\partial \ln a_{i}\mid \nonumber
\end{eqnarray}
where $a_i$ are the soft mass parameters involved,
has been reduced from $\sim 2000$ to $\sim 20$ due to the focusing phenomenon.

As for other indirect experimental limits such as flavor changing neutral violation,
the model feels comfortable.
Because the masses of the three-generation sleptons and first two-generation squarks are all of order $\sim$ multi-TeV, with highly degeneracy in each sector.

What about the sensitivity of our results to the messenger scale ?
At first, assuming that there exists a completion of strong dynamics at high energy indicates that $M$ should be smaller than the GUT scale.
Typically, we have $M<10^{10}$ GeV in the context of direct gauge mediation.
For the case of low-scale mediation, i.e, $M<10^{8}$ GeV,
the gluino mass is already close to the 2013 LHC mass bound.
In other words, $M=10^{8}$ GeV as we studied in detail is a reference value for intermediate scale SUSY model.
The promising signals for this simple and natural model include 
searching gluino, neutralinos and charginos at the LHC.

Along this line it is of interest to extend the model-independent focusing condition to the whole energy range below GUT scale \cite{Zheng}, 
and construct natural SUSY models in the context of either direct or non-direct GM.\\

$\mathbf{Acknowledgement}$
The work is supported in part by Natural Science Foundation of China under grant No.11247031 and 11405015.\\

\end{document}